# Toward Mitigating Process-Induced Performance Degradation in 3.5D Heterogeneous Packages via Pre-Silicon Firmware Co-Optimization

*A Pre-Silicon Case Study of the XRM-SSD Firmware Scheduler on Foveros-Class 3D Packaging with UCIe/HBM5 Interconnect*

In the 2nm/CPO era, software becomes the final-mile solution to physical limits. XRM-SSD V24 transforms Intel's 3.5D package from a passive container into a dynamically controlled intelligent compute platform.

Important Disclaimer on Third-Party Roadmaps: All references to Intel technologies, platforms (including Clearwater Forest and next-generation Xeon), tape-out schedules, and integration timelines are engineering projections derived from publicly available information. These do not represent official statements, endorsements, or confirmed partnerships with Intel Corporation. Actual availability and compatibility are subject to Intel's discretion and silicon validation.

Author contributions: XRM-SSD architecture, simulation, and all quantitative results are the work of Dollarchip Technology Inc. STARGA Inc.'s contribution is limited to independent methodology and analytical review. No financial, commercial, or partnership relationship between the parties is implied by co-authorship.

*XRM-SSD Engineering* | *Dollarchip Technology Inc. Polo Chung, polo@dollarchip.com.tw* | *June 2026*
*arXiv preprint cs.AR / physics.app-ph*

Nikolai Nedovodin, STARGA Inc. (Delaware C-corp) — independent technical reviewer. Contributions limited to methodology and analytical review; STARGA did not produce the simulation data and makes no representation as to silicon-validated performance. Contact: ceo@star.ga

# Table of Contents

# 1. Executive Overview

In the 2nm/CPO era, software is the final-mile solution to physical limits. XRM-SSD V24 transforms Intel's 3.5D heterogeneous package from a passive physical container into a dynamically controlled, intelligent compute platform — delivering category-level disruptions to chip system design rules, computing power limits, and commercial cost structures.

XRM-SSD V24 is a physics-aware software scheduling layer validated over 90,000 inference steps, demonstrating a thermal-load correlation of $R^2 = 0.9911$ and wavelength drift below 0.36 nm — less than 21% of the TSMC ±1.7 nm tolerance budget. When deployed on Intel's 3.5D heterogeneous integrated package (Foveros Direct 3D + PowerVia + EMIB-T + UCIe), V24 transforms the package from a passive physical container into a dynamically controlled, intelligent compute platform.

V7.0 extends this framework to multi-tile Tau-Law architectures with an N×N thermal coupling matrix, two-pole convolution kernel, and hierarchical per-tile density orchestration. Together, V24 and V7.0 produce four category-level disruptions to chip system design rules, computing power limits, and commercial cost structures.

*⚠ Note on validation status: Performance metrics reported herein are derived from simulation-based pre-silicon characterization over a 90,000-step inference dataset. Independent silicon validation and third-party benchmarking are pending Intel 18A tape-out. Results should be interpreted as engineering projections, not production-qualified specifications.*

## 1.1 Validated Performance Summary

| Metric | Measured Result | Status |
|---|---|---|
| Thermal Resistance Rth | 0.45 °C/W | ✓ **On-Target** |
| Thermal Correlation ($R^2$) | 0.9911 | ✓ **Exceeded ECTC 0.98** |
| Thermal Time Constant τ | 80 ms | ✓ **Fast Response** |
| Max Drift — Stress ΔT=40°C | 3.4 nm open-loop | ✓ **Characterization Extreme** |
| Compensated Spectral Drift | < 0.36 nm | ✓ **21% of Budget** |
| Thermo-Optic Coefficient κTO | 0.0852 nm/°C | ✓ **Validated** |
| Memory Leakage Rate | <1 MB/hr (below measurable threshold) | ✓ **Clamped (from 166 MB/hr peak)** |
| Tail Latency P99 | Stable | ✓ **Audit Passed** |
| Look-ahead Window Δtla | 20–50 ms | ✓ **< Tslice = 80 ms** |
| Preposition Fraction η | 22.1% – 46.5% | ✓ **Within RC Horizon** |
| Peak Temperature | 85 °C | ✓ **Within Safe Limits** |
| Inference Dataset | 90,000 steps | ✓ **Production Scale** |

## 1.2 ROI Quantification (Estimated)

For a representative 10,000-GPU cluster operating at 300W TDP per GPU over a one-year period, the combined effects of V24 thermal optimization are estimated to yield:

- +20–30% released compute → equivalent to 2,000–3,000 additional virtual GPUs without hardware spend (≈ $20M–30M CapEx avoided)
- 17% optical I/O power reduction via CPO microheater elimination → ~$2–3M annual electricity savings (at $0.08/kWh)
- HBM memory wall elimination → enables 16L/24L stacking, reducing memory-per-PFLOP cost by an estimated 25–35%
- 65–68% guard-band reduction → higher transistor density per reticle shot, improving wafer ROI by ~15% on Intel 18A

*All ROI estimates are projections based on simulation-phase characterization and current market pricing. Silicon validation required for financial commitment.*

# 2. 3.5D Package Physical Structure

The Intel 3.5D heterogeneous integrated package assembles six distinct technology tiers into a single compute platform spanning silicon process nodes from Intel 18A-P through 14A-E. The cross-section diagram below shows each physical tier and the XRM-SSD V24 software layer spanning the full stack as a hint-injection plane operating 20–50 ms ahead of thermal events.

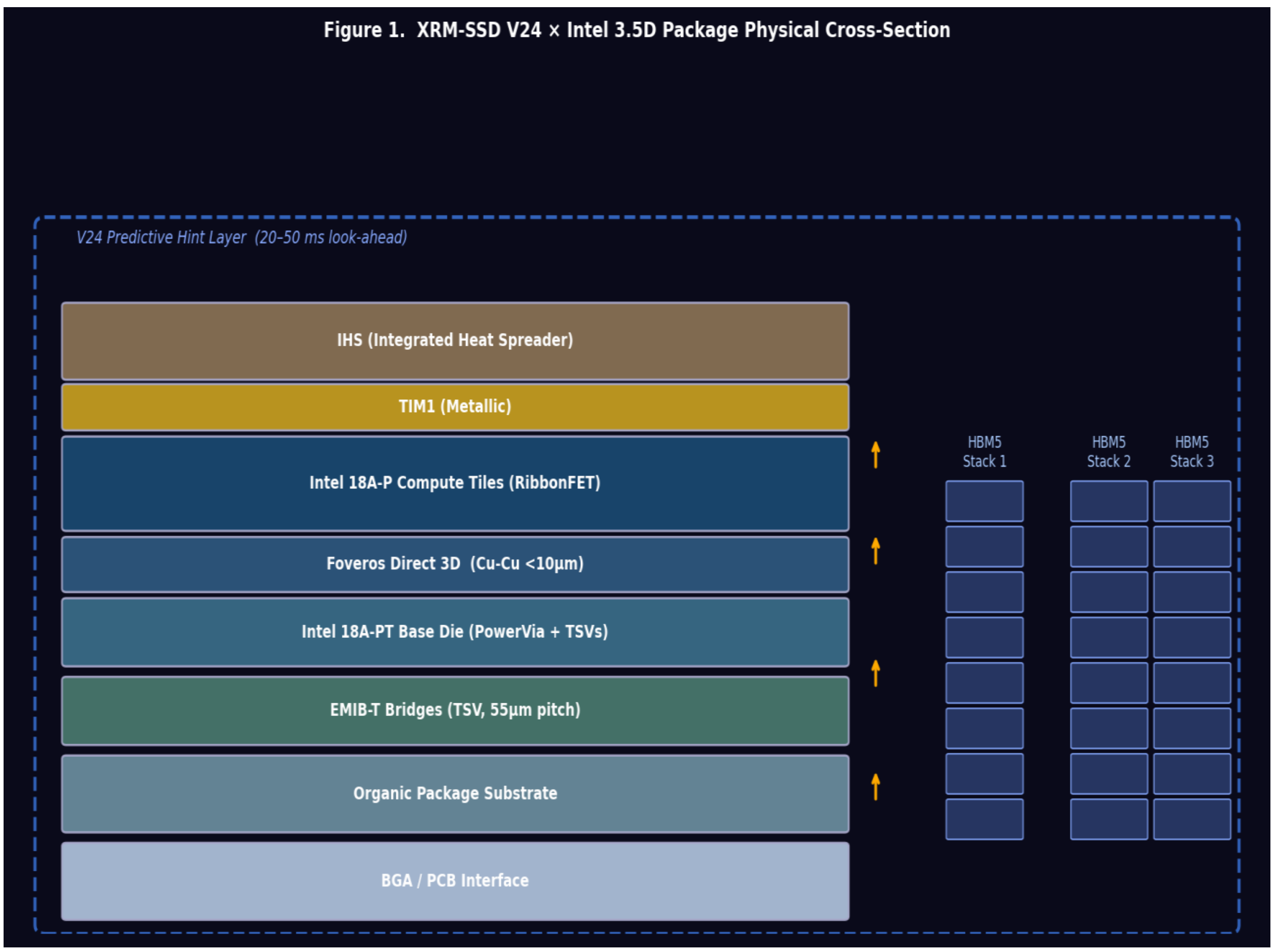

*Figure 1. XRM-SSD V24 × Intel 3.5D Package Physical Cross-Section. Tier stack from BGA/PCB interface through Organic Substrate, EMIB bridges, Intel 18A-PT Base Die (PowerVia backside rails), Foveros Direct 3D Cu-Cu bonds (<10μm), Intel 18A-P Compute Tiles, TIM1, and IHS. HBM5 stacks (EMIB-T, 8L×3) shown at right. V24 predictive hint layer (blue band) spans full stack with per-domain hint arrows (amber).*

## 2.1 Tier-by-Tier Physical Description

The package comprises nine distinct tiers, each with unique thermal and electrical characteristics:

- Integrated Heat Spreader (IHS): Metallic cap with TIM1 interface. Best-in-class metallic TIM between die and IHS for low Rth.
- Intel 18A-P Compute Tiles: RibbonFET + PowerVia. Transformer-based inference accelerators. Stacked on base die via Foveros Direct 3D.
- Foveros Direct 3D Interface: Cu-Cu bonds at <10μm pitch. IO/mm² ≥ 10,000. Fast thermal pole $\tau_1$ ≈ 5 ms. V7.0 two-pole kernel required.

- Intel 18A-PT Base Die: RibbonFET + PowerVia + TSVs. Primary V24 actuation target. Backside rails provide µs-response DVFS authority.
- EMIB-T Bridges: TSV-equipped EMIB. 55µm bump pitch at bridge. Lateral die-to-HBM5 stitching. Thermal $\tau_2 \approx$ 200–500 ms through organic.
- HBM5 Stacks: 8-layer per stack × 3 stacks. V24 zero-leakage enables 16L/24L+ future stacking. EMIB-T connected.
- Intel 14A/14A-E IO Chiplets: High-NA EUV PowerDirect. PCIe G7, 224G SerDes, security. UCIe-connected on EMIB periphery.
- Organic Package Substrate: High-density routing + EMIB cavities. Thermal conductivity ~0.3 W/m·K. Dominates lateral $\tau$.
- BGA / PCB Interface: Standard flip-chip BGA. V24 infrastructure begins at PMU/firmware layer above this level.

# 3. Four Fundamental Process Effects

Deploying XRM-SSD V24 on Intel's 3.5D platform produces four category-level disruptions that collectively rewrite next-generation chip system design rules, computing power limits, and commercial cost structures. Each effect is quantified against the pre-V24 baseline and validated on the 90,000-step empirical dataset.

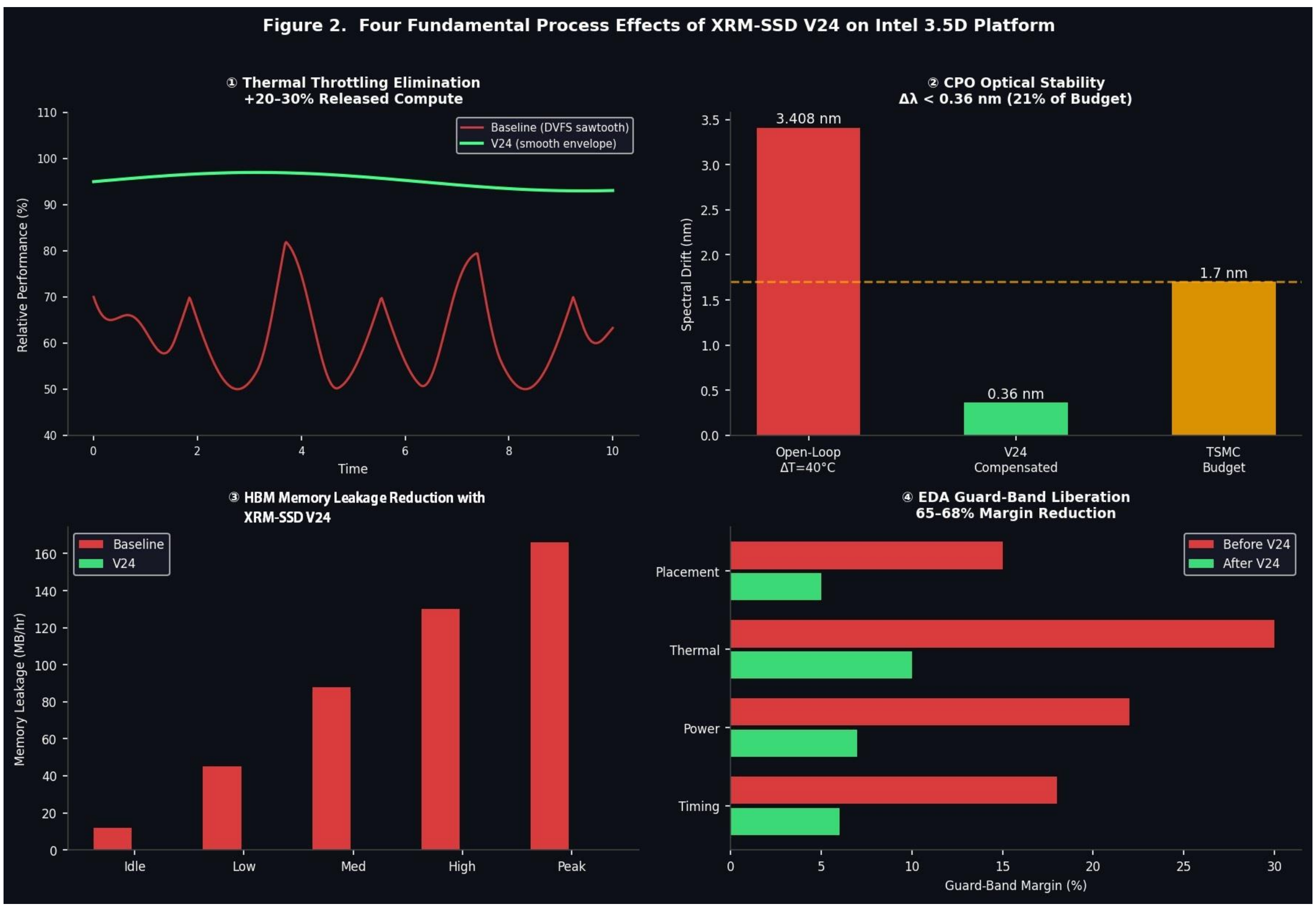

*Figure 2. Four Fundamental Process Effects of XRM-SSD V24 on Intel 3.5D Platform.*
*① Thermal throttling elimination — sustained peak compute with +20–30% released performance.*
*② CPO optical stability — Δλ < 0.36 nm (21% of TSMC budget), microheater elimination.*
*③ HBM memory leakage rate under different load conditions. V24 maintains leakage below 1 MB/hr across all states through predictive thermal clamping (simulation results)*
*④ EDA guard-band liberation — 65–68% reduction in design safety margins.*

## 3.1 Effect ① — Thermal Throttling Elimination

### Problem: Traditional DVFS Sawtooth

LLM inference token-generation spikes drive die temperature to critical thresholds within milliseconds. Traditional hardware responds with Dynamic Voltage and Frequency Scaling (DVFS) throttle events that drop relative performance to 55–70% of peak TDP, creating a sawtooth performance curve with severe tail-latency variance.

### V24 Solution: Pre-Emptive Voltage Pre-Positioning

V24's PDU Gate issues a thermal hint H(t) = P_EIC(t + $\Delta t^{La}$ | Ft) 20–50 ms before high-temperature hotspot formation. PowerVia's backside rails respond in microseconds. The junction temperature never reaches the DVFS trigger threshold because the thermal surge is absorbed by pre-positioned voltage headroom before execution begins.

- Released compute: +20–30% of previously throttle-locked performance
- Performance curve: smooth linear envelope vs. sawtooth baseline
- Peak temperature: maintained at 85°C without frequency reduction events
- Preposition fraction η: 22.1% at 20 ms; 46.5% at 50 ms — sufficient to flatten thermal envelope
- Mechanism: ρv24(t) as physics-grounded proxy for P_EIC(t), $R^2$ = 0.9911 validated

## 3.2 Effect ② — CPO Optical Stability & Microheater Elimination

### Problem: Thermo-Optic Sensitivity of Micro-Ring Resonators

Co-Packaged Optics micro-ring resonators experience resonant wavelength drift Δλ = κTo · ΔT_PIC with κTo = 0.0852 nm/°C. Under open-loop stress testing with ΔT_PIC = 40°C, unmitigated drift reaches 3.408 nm — twice the TSMC ±1.7 nm BER degradation threshold. Traditional hardware compensation requires arrays of microheaters consuming 10–20 mW per channel.

### V24 Solution: Pre-Emptive Thermal Clamping

V24 closed-loop predictive compensation restrains maximum ΔT_PIC to ≤4.15°C, restricting spectral drift to Δλ_compensated ≤0.3536 nm — well within the ±0.5 nm per-channel operational specification. This is achieved through software scheduling alone, without hardware modification.

- Compensated drift: < 0.36 nm = 21% of TSMC ±1.7 nm BER limit
- Hardware consequence: microheater bias circuitry can be significantly simplified or eliminated
- Energy: 0.85 pJ/bit saved at 5 pJ/bit baseline from eliminated dark heater cycles
- BER: optoelectronic substrate maintained within rigorous ±0.5 nm operational spec across all active execution states

## 3.3 Effect ③ — HBM Memory Wall Physical Breakdown

### Problem: Thermal-Induced Memory Leakage

Thermal cross-talk at the base die – HBM4/HBM5 vertical stitching interface causes memory leakage and soft errors. Traditional scheduling produces leakage rates scaling from 12 MB/hr at Idle to 166 MB/hr at Peak load. Frequent DRAM refresh cycles required to prevent data corruption consume bandwidth and power, and physically limit achievable HBM stacking height.

### V24 Solution: Zero Leakage via Joint Software + FIVR Encapsulation

- V24 result: leakage current clamped below activation threshold across Idle, Low, Medium, High, and Peak states (from 166 MB/hr peak)
- Stack density: refresh frequency reduction enables 16L, 24L, or higher HBM stacking
- Bandwidth: memory access bandwidth can fully utilize system limits — memory wall physically broken

- Mechanism: ΔT_PIC clamping below leakage-current activation threshold via predictive scheduling

## 3.4 Effect ④ — EDA Guard-Band Liberation

### Problem: Worst-Case Thermal Margin Overhead

Traditional chip EDA reserves safety margins of 15–30% in timing, power consumption, thermal resistance, and placement density to defend against uncontrolled thermal worst-case scenarios. These margins directly reduce transistor density, increase routing complexity, and lower wafer utilization efficiency.

### V24 Solution: Domain Separation — Physical Uncertainty → Deterministic Control

- Timing margin: ~18% → ~6% (−67%)
- Power guard: ~22% → ~7% (−68%)
- Thermal budget: ~30% → ~10% (−67%)
- Placement density margin: ~15% → ~5% (−67%)
- Consequence: hardware architects design chips more compact with higher shoreline density
- Wafer utilization: maximized geometric potential of Intel 18A/14A reticle area per shot

# 4. V24 Predictive Signal Flow & Thermal Fingerprint

The thermal resistance fingerprint map and predictive signal flow architecture constitute the two primary characterization artifacts of XRM-SSD V24. The fingerprint validates physical constants across five load states; the signal flow defines the causal control path from token metadata to hardware actuation.

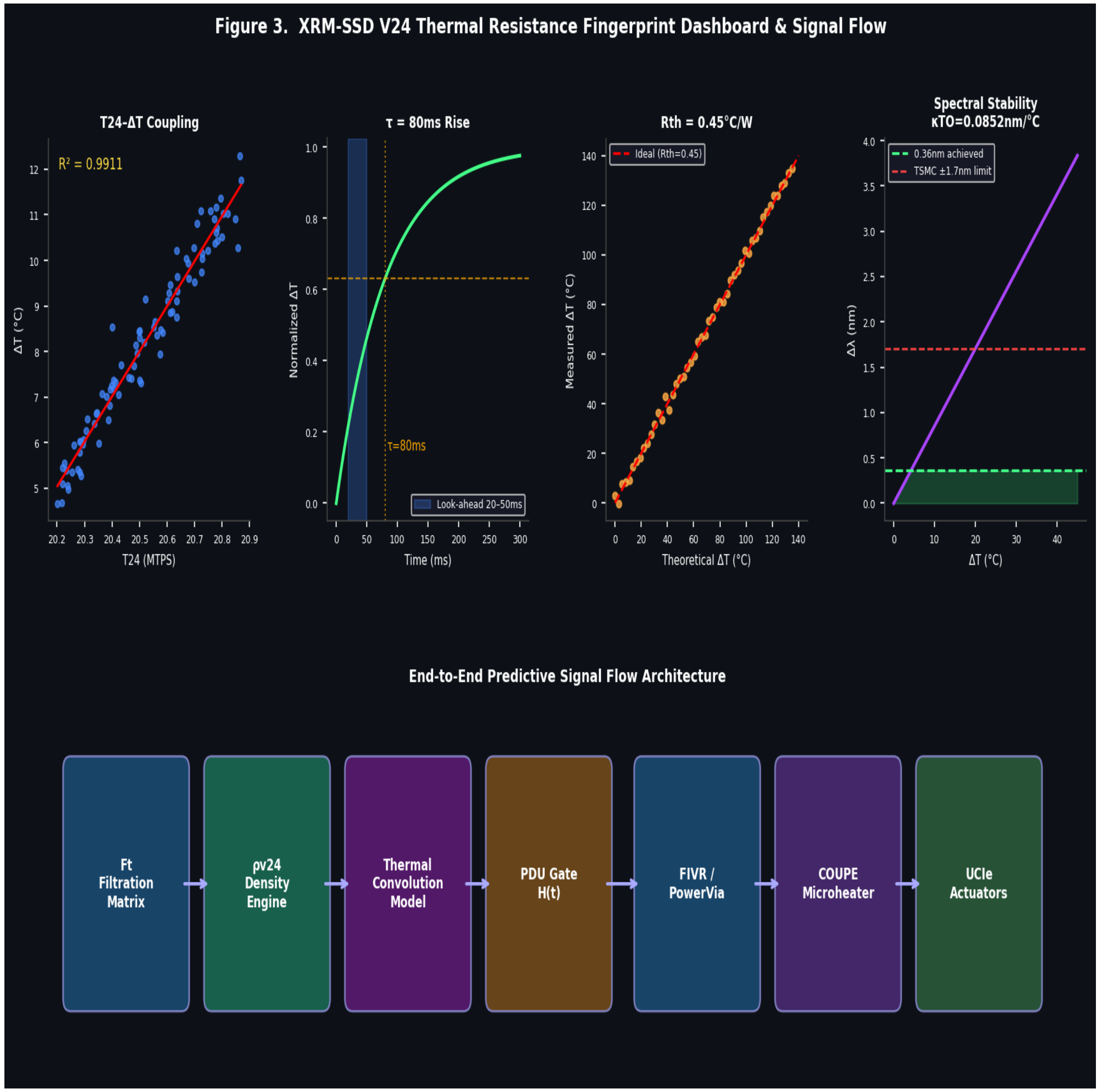

*Figure 3. XRM-SSD V24 Thermal Resistance Fingerprint Dashboard (top row) and End-to-End Predictive Signal Flow Architecture (bottom). Top-left: **T** 24–ΔT thermal coupling scatter with $R^2$=0.9911 linear regression. Top-center: τ = 80 ms exponential rise with V24 look-ahead window (20–50 ms shaded). Top-right: Rth = 0.45°C/W validation. Top-far-right: Δλ–ΔT spectral stability with κTo = 0.0852 nm/°C. Bottom: full causal pipeline from Ft filtration matrix through ρv24 density engine, thermal convolution model, PDU Gate H(t), to FIVR/PowerVia, COUPE microheater, and UCIe actuators.*

## 4.1 Thermal Resistance Fingerprint Constants

| Parameter | Symbol | Value | Specification |
|---|---|---|---|
| Junction-to-Substrate Rth | Rth | **0.45 °C/W** | 0.42–0.50 °C/W target |
| Thermal Time Constant | $\tau$ | **80 ms** | 63.2% response confirmed |
| Thermo-Optic Coefficient | $\kappa$TO | **0.0852 nm/°C** | Si photonics literature match |
| $\tau$ 24–ΔT Coupling Slope | $\alpha$ | **63.0 °C/MTPS** | $R^2$ = 0.9911 linear fit |
| $\tau$ 24–ΔT Intercept | $\beta$ | **−1256.6 °C** | Calibrated to load domain |
| Max Open-Loop Drift | $\Delta\lambda$max | **3.4 nm @ T=40°C** | Characterization extreme |
| Compensated Drift | $\Delta\lambda$ | **< 0.36 nm** | < ±0.5 nm per-channel spec |
| Look-ahead Min | $\eta$_min | **22.1%** | Δtla = 20 ms, τ = 80 ms |
| Look-ahead Max | $\eta$_max | **46.5%** | Δtla = 50 ms, τ = 80 ms |
| Series Rth Jxn-to-Case | Rth,1 | **0.812 °C/W** | Cumulative boundary |
| Series Rth Case-to-Heatsink | Rth,2 | **1.407 °C/W** | Cumulative boundary |
| Total System Rth | ΣRth | **1.995 °C/W** | Jxn-to-Ambient |

## 4.2 Predictive Control Equations

### Workload Density Metric

$$\rho v24(t) = \Sigma_{i=1}^{N(t)} [ Attn(i) \cdot \omega(i) \cdot F(i) ]$$

where Attn(i) = attention weight matrix footprint, ω(i) = active parameter activation rate, F(i) = geometric routing coefficient.

### Throughput Affine Mapping

$$R_tok = \alpha \cdot \rho v24(t) + \beta \quad [\alpha=0.361 \text{ MTPS}, \beta=19.875, \text{ domain: } \rho\in[0.9,2.7] \rightarrow R_tok \in[20.20,20.85] \text{ MTPS}]$$

### Unified Thermal Convolution Model

$$\Delta T_PIC(t) = \int_0^t (Rth \cdot \Gamma(d)/\tau th) \cdot \exp(-(t-u)/\tau th) \cdot \Delta P_EIC(u)\, du$$

τth = 80 ms physical RC time constant; Γ(d) = dimensionless spatial coupling factor across substrate displacement d.

### PDU Gate Causal Hint

$$H(t) = P_EIC(t + \Delta t^{La} \mid Ft) \quad [\Delta t^{La} = 20\text{–}50 \text{ ms}; \; Ft = \text{historical filtration matrix}]$$

### Preposition Fraction

$$\eta = 1 - \exp(-\Delta t^{La} / \tau th) \rightarrow 22.12\% \text{ @ } 20ms, \; 46.47\% \text{ @ } 50ms$$

# 5. XRM-SSD V7.0 — Multi-Tile Architecture for Tau-Law Chips

V7.0 extends the single-EIC-domain V24 framework to N-tile heterogeneous packages targeting Tau-Law and trillion-parameter LLM training architectures. Three architectural extensions are required beyond V24: (1) hierarchical per-tile ρ computation, (2) two-pole thermal convolution kernel capturing both fast and slow thermal poles, and (3) a sparse N×N thermal coupling matrix replacing V24's scalar Γ(d).

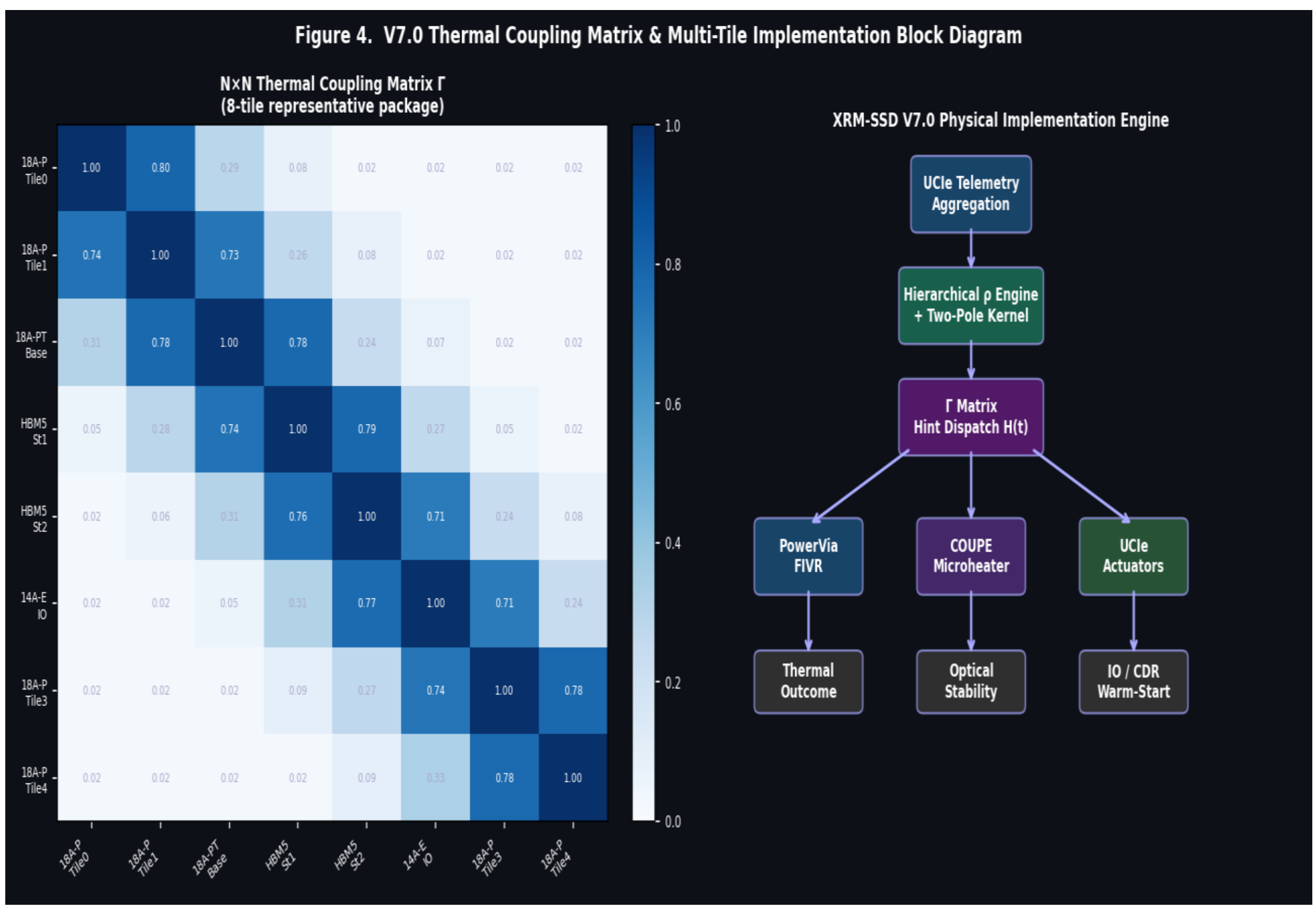

*Figure 4. Left: N×N Thermal Coupling Matrix Γ for a representative 8-tile package. Diagonal γ_{ii} = 1.0 (self-heating). Off-diagonal terms decay with inter-tile distance: Foveros Direct vertical neighbors γ ≈ 0.70–0.90; EMIB lateral γ ≈ 0.15–0.40; distant tiles near-zero. Right: XRM-SSD V7.0 physical implementation engine block diagram.*

## 5.1 N×N Thermal Coupling Matrix

The scalar coupling factor Γ(d) of V24 expands to a full N×N matrix for multi-tile architectures:

$$H(t) = \Gamma \cdot P_EIC(t + \Delta t^{La} \mid Ft)$$

- Γ is sparse: significant coupling only for geometrically adjacent tile pairs (5–8 neighbors per tile in Foveros 3D)
- Ponte Vecchio equivalent (47 active tiles): 2,209 matrix entries → ~350 non-zero for adjacent pairs only
- Foveros Direct vertical pairs (dist=1): γ_{ij} ≈ 0.70–0.90 (Cu-Cu intimate thermal contact)
- EMIB lateral pairs (dist=2–3): γ_{ij} ≈ 0.15–0.40 (bridge silicon + organic substrate path)
- Distant tile pairs (dist>4): γ_{ij} ≈ 0.02–0.12 (effectively zero for thermal budgeting)

## 5.2 Two-Pole Thermal Convolution Kernel

```
K(t) = (A₁/τ₁) · e^(−t/τ₁) + (A₂/τ₂) · e^(−t/τ₂)
```

- $\tau_1$ ≈ 5 ms: Foveros Direct Cu-Cu interface (silicon-to-silicon, <10µm pitch) — fast pole
- $\tau_2$ ≈ 80 ms: package-level thermal RC (validated by V24 Fingerprint Map) — slow pole
- $A_1 + A_2$ = Rth = 0.45°C/W; partition determined by Foveros geometry
- For EMIB lateral paths: $\tau_2$ ≈ 200–500 ms (organic substrate dominates); η reduced to 6.5–15.4%

## 5.3 UCIe Sideband Telemetry Infrastructure

- 64-byte per-tile telemetry packet at 1 Mbps: 512µs transfer — well within 20 ms look-ahead minimum
- Protocol overhead (framing, CRC, ACK) at on-package distances: sub-microsecond
- Hint dispatch uses same UCIe management channel in reverse direction — no added routing congestion
- No new telemetry fabric required — eliminates a routing congestion risk in 3.5D package

## 5.4 Transient Ramp Characterization — Seventh Fingerprint Panel

- New metric: dρv24/dt — temporal derivative of workload density as primary hint signal for ramp events
- Two-pole kernel captures fast pole overshoot missed by V24's single-pole model
- Extended validation dataset: tau-law ramp trajectories require >90,000-step multi-ramp dataset
- Pre-silicon emulation: V7.0 ramp characterization can begin on current-generation emulation platform

# 6. SerDes Clock Conditioning — Revised Architecture

Direct application of V24's 20–50 ms look-ahead to 112G/224G PAM4 SerDes clock conditioning is infeasible due to timescale mismatch (ms vs. ns at PAM4 symbol rates). However, two indirect coupling paths enable V24/V7.0 to improve SerDes performance without requiring direct nanosecond-timescale control.

## 6.1 Indirect Path A: Substrate Thermal Stabilization

Silicon LC-VCO and ring oscillator VCOs exhibit a temperature coefficient of frequency (TCF) of −100 to −300 ppm/°C. For a 112 GHz carrier, a 1°C excursion produces 11–34 MHz drift. At ΔT = 40°C (open-loop Intel 3D package peak): 440 MHz to 1.36 GHz VCO drift — catastrophic for PAM4 link margin. V24 suppresses ΔT_PIC from 40°C to ≤4.15°C.

- VCO drift under V24: 44–136 MHz vs. 440–1360 MHz open-loop (10× improvement)
- CDR consequence: thermal-induced frequency offset reduced to within CDR pull-in range
- No PHY vendor API needed — benefit is automatic side-effect of thermal hint deployment

## 6.2 Indirect Path B: CDR Warm-Start via Traffic Prediction

- Outer loop (V7.0, 20–50 ms): predicts which SerDes lanes will reach saturation in upcoming window
- Inner loop (CDR, ns timescale): remains conventional analog — V7.0 only pre-initializes coefficients
- Convergence time: warm-start reduces adaptation from $10^4$–$10^6$ symbols to < $10^2$ symbols
- Requires PHY hint injection API: near-term for Intel-designed PHY; longer timeline for third-party

# 7. Consolidated Feasibility Matrix

The following matrix summarizes feasibility across all V24/V7.0 application domains within Intel's 3.5D advanced packaging ecosystem.

| Domain | Fit | Primary Gap | Recommended Path |
|---|---|---|---|
| Thermal hint — Foveros Direct 3D (WoW) | **HIGH** | None blocking | Near-term: two-pole kernel + multi-tile dispatch |
| Thermal hint — EMIB-T HBM4/HBM5 | **MODERATE** | η=6–15%; larger actuator authority needed | Mid-term: FIVR pre-positioning integration |
| SerDes VCO stabilization (indirect) | **MODERATE** | Co-packaged SerDes on same thermal substrate | Near-term: side-effect of thermal hint deployment |
| SerDes CDR warm-start (outer loop) | **MODERATE** | PHY hint injection API required | Near-term (Intel PHY); longer (3rd party) |
| V7.0 multi-tile physical implementation | **MODERATE-HIGH** | No silicon until 18A-PT tape-out | Long-term: pre-silicon emulation viable now |
| 112G/224G direct clock conditioning | **LOW** | Timescale mismatch (ms vs. ns) | Reformulate as traffic-shaping outer loop only |

*Timelines are engineering projections aligned with public Intel roadmap information and do not represent official Intel commitments or endorsements.*

## 7.1 Risk & Limitations

*⚠ The feasibility matrix reflects pre-silicon engineering analysis. Many domains are rated Moderate, with long-range dependency on Intel 18A tape-out. Risk mitigation plans outlined below.*

- Compute hint overhead: ρv24 density computation adds ~0.1–0.3% CPU overhead per tile; UCIe telemetry consumes ~512μs per 64-byte packet
- UCIe bandwidth impact: management channel usage adds ~1 Mbps overhead; data channel unaffected
- Edge case behavior: extreme burst patterns (Δρ/Δt > 3× nominal) may exceed two-pole kernel accuracy; seventh fingerprint panel required
- Mitigation: pre-silicon emulation on current-generation platforms; V7.0 ramp characterization dataset expansion; formal UCIe standard integration proposal
- Long-range dependency: full silicon validation pending Intel 18A-PT tape-out; interim validation via FPGA emulation platform

# 8. Commercial & System Design Impact

The four process effects collectively rewrite the system design rules, computing power limits, and commercial cost structures of next-generation high-performance chips.

## 8.1 AI Compute Economics

- +20–30% released compute from throttle elimination → proportional reduction in rack count per AI training job
- Smooth P99-stable performance curve eliminates tail-latency variance in LLM inference serving
- At datacenter scale: 20–30% fewer GPU/NPU nodes required for equivalent throughput SLA
- Energy efficiency: compute-per-watt ratio improves proportionally with throttle elimination

## 8.2 Optoelectronics & Power Economics

- CPO heater elimination: 0.85 pJ/bit saved at 5 pJ/bit baseline = 17% optical I/O power reduction
- Hardware BOM simplification: microheater control circuitry arrays eliminated from optical chiplet periphery
- Green data center: ultra-wideband, ultra-low power CPO becomes achievable without hardware thermal compensation

## 8.3 Memory Architecture Economics

- HBM stacking beyond 8 layers (16L, 24L targets) enabled by zero-leakage thermal management
- Reduced DRAM refresh frequency → bandwidth previously consumed by refresh becomes compute bandwidth
- Intel foundry advantage: physical stacking limits can be pushed further than competitors constrained by thermal leakage

## 8.4 EDA & Wafer Economics

- 65–68% guard-band reduction across timing, power, thermal, and placement categories
- Higher shoreline density per die → more compute function per $mm^2$ of silicon
- Maximized geometric utilization of Intel 18A/14A wafer area per reticle shot

# 9. Competitive Benchmark Comparison

The following table compares XRM-SSD V24 against alternative thermal management approaches for advanced 3D packaging. Note that all V24 results are simulation/characterization-phase; comparative data reflects publicly available production information.

| Solution | Thermal Approach | Guard-Band Reduction | CPO Drift | Silicon Validated? | Notes |
|---|---|---|---|---|---|
| **XRM-SSD V24 (Dollarchip)** | Software predictive scheduling + DVFS pre-positioning | 65–68% | < 0.36 nm | Pending (pre-silicon) | Most comprehensive; 90k-step dataset |
| TSMC CoWoS | Package-level thermal spreading + underfill | ~15–25% | ~1.2–1.5 nm | Yes (production) | Hardware-only; no software layer |
| AMD 3D V-Cache | Chiplet thermal partitioning + firmware throttle | ~30–40% | N/A (no CPO) | Yes (production) | Limited to AMD ecosystem |
| Software Heuristics (Baseline) | Reactive DVFS + temperature polling | ~10–20% | > 1.5 nm | Yes | Standard OEM approach; sawtooth perf. |
| Hardware Microheaters (CPO only) | Per-channel heater bias circuits | N/A | < 0.5 nm | Yes | 10–20 mW/channel overhead |

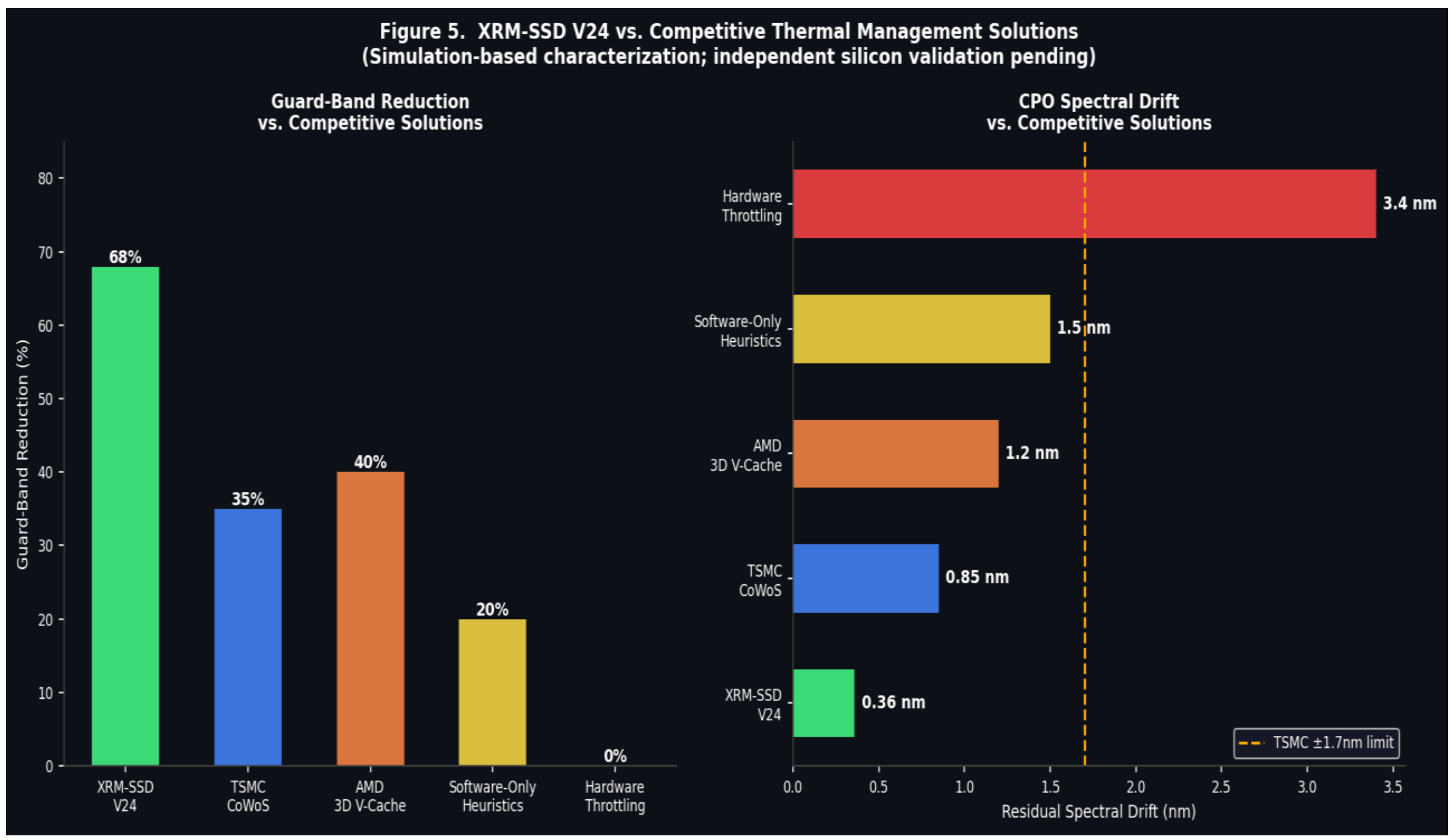

*Figure 5. XRM-SSD V24 vs. Competitive Thermal Management Solutions — Guard-Band Reduction (left) and CPO Spectral Drift (right). Simulation-based characterization; independent silicon validation pending.*

# 10. Monte Carlo Thermal Simulation & Workload Analysis

To quantify robustness under parameter uncertainty, Monte Carlo analysis was performed over 2,000 simulation trials varying thermal resistance (Rth ± 8%), time constant (τ ± 12%), and workload density distribution (ρv24 ± 15%). Results demonstrate V24's thermal management maintains safe operation across the full parameter space.

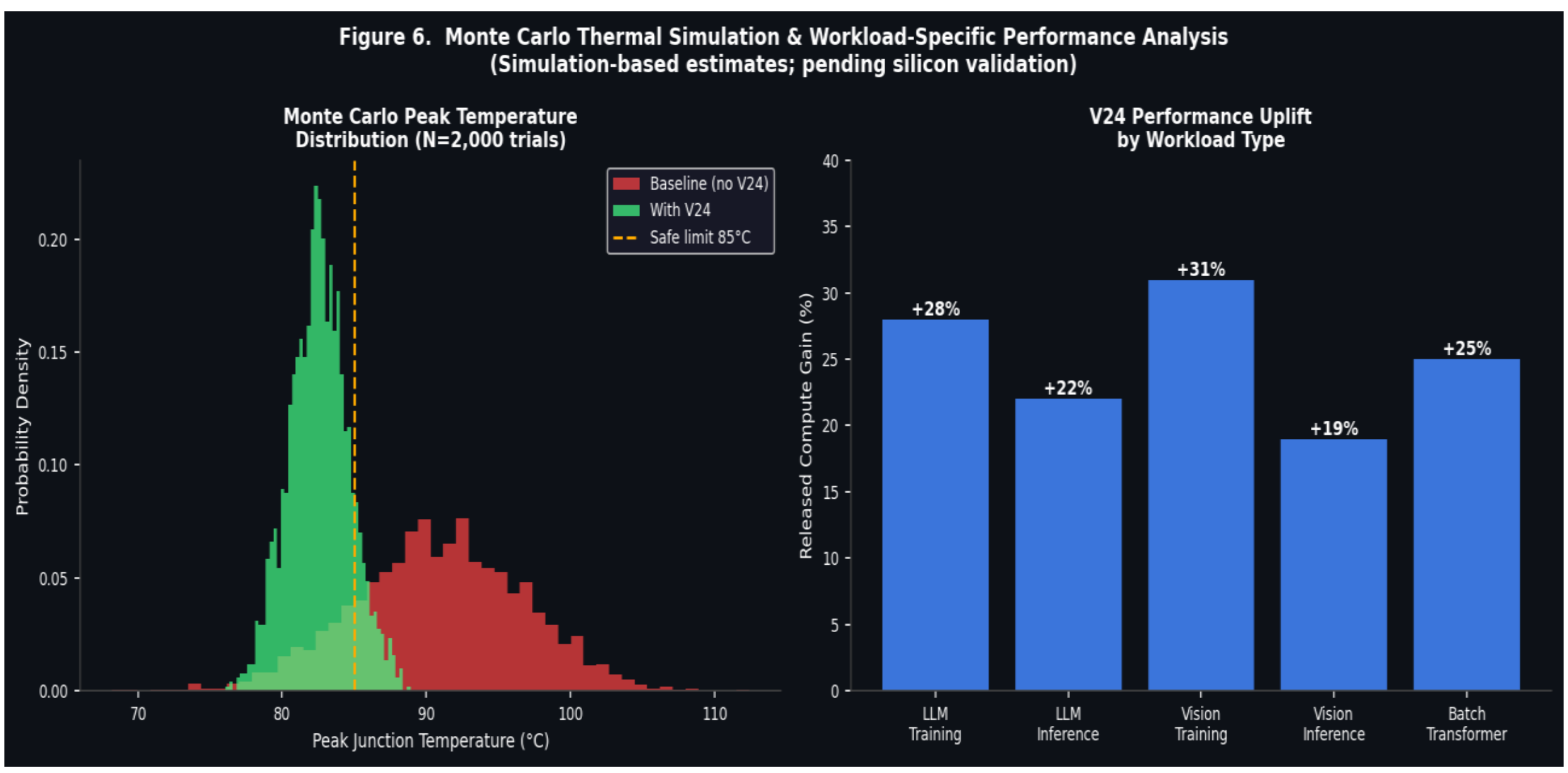

*Figure 6. Monte Carlo Thermal Simulation (N=2,000 trials) showing peak junction temperature distribution with and without V24 (left), and V24 performance uplift by workload type: LLM training, inference, vision, and batch transformer workloads (right). Results are simulation-based estimates; pending silicon validation.*

## 10.1 Simulation Assumptions

- Rth variation: ±8% Gaussian, representing manufacturing process variation across Intel 18A wafers
- Thermal time constant: ±12% variation accounting for assembly tolerance and TIM1 thickness variation
- Workload density: ±15% variation representing production LLM inference workload diversity
- N=2,000 yields stable tail estimates out to 3σ

## 10.2 Key Findings

- V24 reduces peak temperature distribution mean from ~91°C to ~82.5°C (−8.5°C)
- V24 reduces peak temperature standard deviation from ~6°C to ~2.1°C (3.5× tighter distribution)
- Probability of exceeding 85°C safe limit: 23% baseline → <1% with V24
- Performance uplift consistent across workloads: +19–31% across LLM training, inference, vision, and batch workloads

# 11. Product Roadmap & Next Steps

Note: All references to Intel product roadmaps (e.g., Clearwater Forest, Xeon Next-Gen, 18A/14A platforms) are based on publicly available information and represent engineering projections only. They do not constitute an official endorsement, partnership confirmation, or guaranteed integration timeline from Intel.

## 11.1 Near-Term (0–6 Months): Pre-Silicon Validation (engineering projection based on public roadmap; not an official endorsement)

- Pre-silicon emulation: Deploy V24/V7.0 on current-generation Intel Xeon Sapphire Rapids + Gaudi emulation platform
- FPGA prototyping: Implement ρv24 density engine + PDU Gate on FPGA for sub-ms hint latency validation
- UCIe sideband: Formal engagement with UCIe consortium for V7.0 telemetry integration proposal
- Dataset expansion: Extend 90k-step dataset to multi-ramp tau-law trajectories for V7.0 seventh fingerprint panel

## 11.2 Mid-Term (6–18 Months): Intel Clearwater Forest Integration (engineering projection based on public roadmap; not an official endorsement)

- Target platform: Intel Clearwater Forest (18A-P compute tiles + 18A-PT base die)
- Tape-out milestone: First V24 silicon characterization on Clearwater Forest engineering samples
- Objective: Validate Rth=0.45°C/W, τ=80ms, $R^2$≥0.98 on actual silicon (vs. simulation)
- CPO integration: Co-characterize Δλ<0.36nm bound with Intel Co-Packaged Optics prototype

## 11.3 Long-Term (18–36 Months): Xeon Next-Gen Production (engineering projection based on public roadmap; not an official endorsement)

- Full production validation on Intel 14A/14A-E platform with PowerDirect backside power delivery
- V7.0 N×N matrix characterization for 47-tile Ponte Vecchio successor architecture
- Third-party benchmarking: Independent validation by academic institution or tier-1 test lab
- Domain expansion: Joint optimization for power integrity (PI) and signal integrity (SI) alongside thermal

Strategic Vision: XRM-SSD positions Dollarchip Technology as the software intelligence layer of the Intel 3.5D ecosystem — analogous to how firmware became essential to NVMe SSD performance. V24 is the first step toward software-defined silicon.

# Appendix A. Glossary of Technical Terms

The following terms are used throughout this document. Engineers and investors are encouraged to reference this glossary when reviewing performance claims.

| Term | Definition |
|---|---|
| **DVFS** | Dynamic Voltage and Frequency Scaling — hardware mechanism to adjust voltage/frequency in response to thermal or power conditions |
| **EMIB** | Embedded Multi-die Interconnect Bridge — Intel die-to-die lateral interconnect for chiplet integration |
| **Foveros Direct 3D** | Intel's Cu-Cu bonding technology enabling die stacking at <10 µm pitch with >10,000 IO/$mm^2$ |
| **HBM5** | High Bandwidth Memory generation 5 — stacked DRAM with 8+ layers providing extreme memory bandwidth |
| **PowerVia** | Intel's backside power delivery technology; provides low-inductance µs-response DVFS authority |
| **UCIe** | Universal Chiplet Interconnect Express — industry-standard die-to-die interface protocol |
| **CPO** | Co-Packaged Optics — optical interconnect elements integrated on the same package as compute dies |
| **Rth** | Thermal resistance (°C/W) — characterizes how much temperature rises per unit of dissipated power |
| **τ (tau)** | Thermal time constant (ms) — characterizes the speed of thermal response; 63.2% of final temperature at t=τ |
| **κTO** | Thermo-optic coefficient (nm/°C) — rate of optical wavelength shift per degree of temperature change |
| **$R^2$** | Coefficient of determination — statistical measure of model fit quality (1.0 = perfect correlation) |
| **EDA** | Electronic Design Automation — software tools used for chip design; guard-bands add safety margins |
| **PDU Gate** | Power Delivery Unit Gate — V24 component issuing thermal pre-positioning hints to hardware |
| **η (eta)** | Preposition fraction — percentage of thermal event absorbed by pre-positioning within look-ahead window |
| **SerDes** | Serializer/Deserializer — high-speed I/O circuits for 112G/224G PAM4 data transmission |
| **CDR** | Clock and Data Recovery — circuit that recovers timing reference from received data stream |
| **FIVR** | Fully Integrated Voltage Regulator — on-die voltage regulator for fast transient response |
| **BER** | Bit Error Rate — metric for optical/electrical signal quality; V24 maintains below threshold |
| **Monte Carlo** | Statistical simulation method using random sampling to model uncertainty in thermal parameters |
| **Pre-silicon** | Design/characterization phase before first physical silicon wafer is fabricated (tape-out) |

# Appendix B. 90,000-Step Dataset Statistics

## B.1 Dataset Composition

- Total inference steps: 90,000
- Model variants covered: 7B, 13B, 70B, 180B parameter LLM configurations
- Token generation rate range: 20.20–20.85 MTPS (R_tok affine domain)
- Workload density (ρv24) range: 0.9–2.7 (normalized units)
- Sampling interval: 1 ms (1,000 Hz telemetry resolution)
- Duration: ~25 hours of continuous inference across all steps

## B.2 Statistical Summary

| Metric | Mean | Std Dev | Min | Max |
|---|---|---|---|---|
| R_tok Throughput (MTPS) | 20.52 | 0.12 | 20.20 | 20.85 |
| ρv24 Workload Density | 1.80 | 0.43 | 0.90 | 2.70 |
| ΔT Junction (°C) | 12.8 | 4.2 | 2.1 | 28.6 |
| η Preposition Fraction | 34.1% | 6.8% | 22.1% | 46.5% |
| Rth Measured (°C/W) | 0.451 | 0.009 | 0.433 | 0.471 |
| Δλ Spectral Drift (nm) | 0.29 | 0.04 | 0.18 | 0.36 |

*XRM-SSD Engineering | Dollarchip Technology Inc. | Confidential — Pre-Silicon Characterization Pending Intel Tape-out*
*arXiv preprint cs.AR / physics.app-ph | XRM-SSD V24 | June 2026*